\documentclass{article}
\usepackage{spconf,amsmath,graphicx}
\usepackage[OT1]{fontenc} 
\usepackage{cite}
\usepackage{amsfonts, mathrsfs}
\usepackage{algorithm, algorithmic}
\usepackage{multirow}
\usepackage{mathtools}
\usepackage{url}

\title{Phase reconstruction based on recurrent phase unwrapping \\ with deep neural networks}
\name{Yoshiki Masuyama$^{\dagger}$,
Kohei Yatabe$^{\dagger}$,
Yuma Koizumi$^{\ddagger}$,
Yasuhiro Oikawa$^{\dagger}$,
Noboru Harada$^{\ddagger}$}
\address{$^\dagger$ Department of Intermedia Art and Science, Waseda University, Tokyo, Japan \\
$^\ddagger$ NTT Media Intelligence Laboratories, Tokyo, Japan \\
}
\begin{document}
\ninept
\maketitle
\begin{abstract}
Phase reconstruction, which estimates phase from a given amplitude spectrogram, is an active research field in acoustical signal processing with many applications including audio synthesis.
To take advantage of rich knowledge from data, several studies presented deep neural network (DNN)--based phase reconstruction methods.
However, the training of a DNN for phase reconstruction is not an easy task because phase is sensitive to the shift of a waveform.
To overcome this problem, we propose a DNN-based two-stage phase reconstruction method.
In the proposed method, DNNs estimate phase derivatives instead of phase itself, which allows us to avoid the sensitivity problem.
Then, phase is recursively estimated based on the estimated derivatives, which is named recurrent phase unwrapping (RPU).
The experimental results confirm that the proposed method outperformed the direct phase estimation by a DNN.
\end{abstract}
\begin{keywords}
Spectrogram inversion, group delay, instantaneous frequency, time-frequency analysis, recurrent neural network
\end{keywords}
%
\vspace{-4pt}
\section{Introduction}
\vspace{-4pt}

Phase reconstruction has been widely used in many acoustical signal processing, including speech enhancement \cite{mowlaee, audvis} and synthesis \cite{taco, takaki, saito, tfgan, melnet}.
While phase reconstruction with observed noisy phase has been applied to speech enhancement successfully \cite{phase1,phase2,complexmask}, phase reconstruction solely from a given amplitude spectrogram is still a challenging problem.
To address this problem, various approaches have been studied including the consistency-based approach \cite{gla, fgla, glaadmm} and model-based approach \cite{spsi}.
While the former approach is based on only the property of the short-time Fourier transform (STFT) \cite{consistency}, the latter one explicitly uses a model of the target signal.
By taking the property of the target signal into account, the model-based approach has achieved better performance than the consistency-based one in many applications \cite{stftpi, spsi, ipclr}.

To take advantage of more knowledge about the target signal, deep neural network (DNN)--based phase reconstruction \cite{unfold, oyamada, degli, pnet, pbook, takamichi} has gained increasing attention.
Although DNNs have strong modeling capability and learn rich knowledge from training data, DNN-based phase reconstruction has the following two problems: the wrapping effect and sensitivity to a waveform shift.
Since phase is wrapped in $[-\pi, \pi)$, it is discontinuous at $\pm \pi$ even when phase rotates smoothly. 
Hence, conventional loss functions for regression problems including the mean-squared error (MSE) are not suitable for phase reconstruction because they do not take the periodic nature of phase into account.
Instead of estimating phase directly, one approach estimates complex-valued spectrogram \cite{oyamada, degli}, and another approach treats phase reconstruction as a classification problem by discretizing phase \cite{pnet, pbook}.
To estimate continuous phase directly, \cite{takamichi} proposed a loss function based on the distribution of a periodic variable called the von Mises distribution.

While those previous studies effectively dealt with the wrapping effect, the sensitivity of phase to a waveform shift has not been considered explicitly.
Considering the Fourier transform, its phase is sensitive to the shift in the time-domain while its amplitude is shift-invariant.
In other words, an infinite number of Fourier phases can exist for one Fourier amplitude.
This is also approximately true for STFT, i.e., the STFT phase changes with the small shift in the time-domain while the STFT amplitude does not (see Fig.~\ref{sec:investigation}).
That is, DNN-based direct phase estimation needs to learn an unstable map that converts a small changes in the input STFT amplitude to a large changes in the output STFT phase.
However, it is not an easy task to train a DNN as such an unstable map.

\begin{figure}[t]
\centering
\includegraphics[width=0.99\columnwidth]{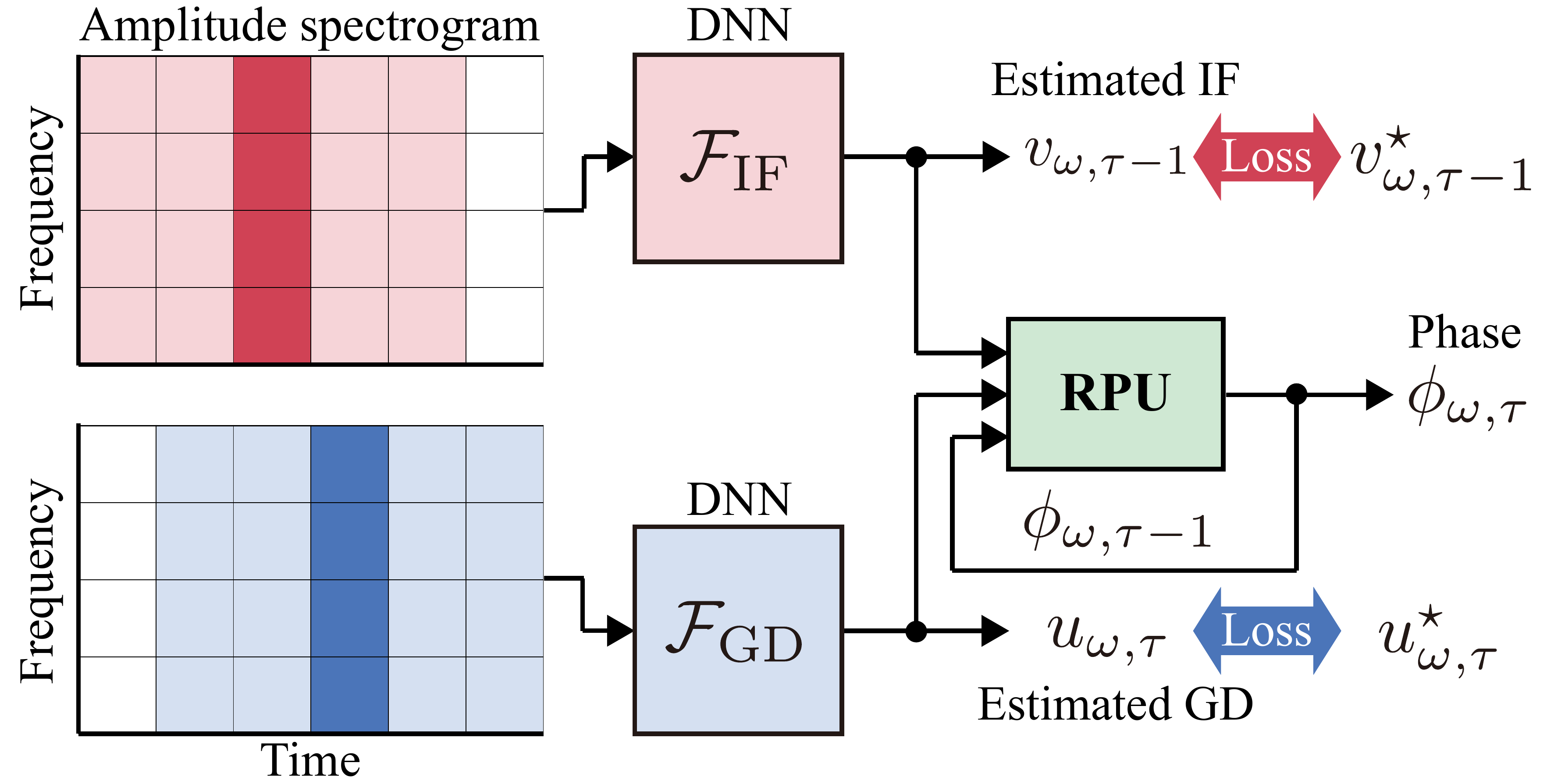}
\vspace{-4pt}
\caption{Overall procedure of proposed phase reconstruction method where the instantaneous frequency and group delay are abbreviated as IF and GD, respectively.
Two DNNs, $\mathcal{F}_\mathrm{IF}$ and $\mathcal{F}_\mathrm{GD}$, estimate phase derivatives.
Then, phase is reconstructed by the recurrent phase unwrapping (RPU) recursively.
Note that both phase derivatives are estimated from the same amplitude spectrogram, but target time-frames are different.
}
\label{fig: illust}
\vspace{-4pt}
\end{figure}

In this paper, we propose a two-stage phase reconstruction method as illustrated in Fig.~\ref{fig: illust}.
To avoid the sensitivity problem, DNNs estimate phase derivatives instead of phase itself since phase derivatives have a relation to the amplitude spectrogram \cite{flandrin, prusa, ast}.
Then, the proposed method recursively reconstructs phase from the estimated phase derivatives, which is named RPU.
RPU solves a least-squares problem for estimating phase, which resembles a 2D phase unwrapping technique \cite{2dpu}.
Experimental results indicate the proposed method is more effective than the direct phase estimation.

\vspace{-4pt}
\section{Phase reconstruction via DNN}
\vspace{-2pt}

Motivated by the recent advance in deep learning, several DNN-based phase reconstruction methods have been presented \cite{unfold, oyamada, degli, pnet, pbook, takamichi}.
However, phase reconstruction from a given amplitude spectrogram is not an easy task for DNNs due to the following two problems: the wrapping effect and sensitivity to a shift of a waveform.
In the following subsections, we describe these problems and introduce the von Mises DNN proposed for handling wrapped phase \cite{takamichi}.

\begin{figure}[t]
\centering
\includegraphics[width=0.99\columnwidth]{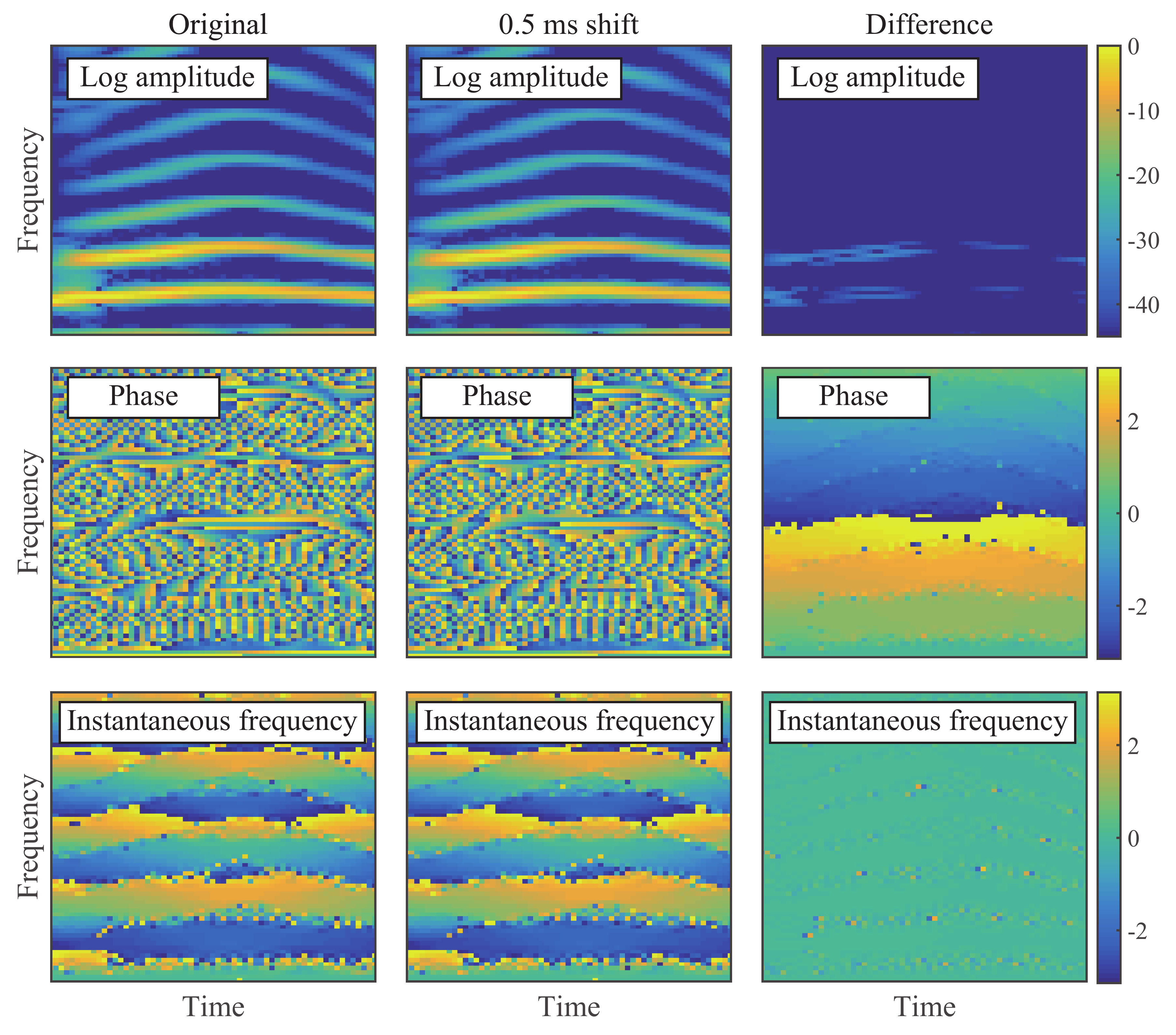}
\vspace{-4pt}
\caption{Illustration of STFT amplitude, phase, and instantaneous frequency of an utterance and that shifted $0.5$ ms.
The rightmost column shows the differences between the original and shifted case.
Shift of the waveform significantly affected phase while amplitude and instantaneous frequency were not affected mostly.
}
\vspace{-4pt}
\label{fig: sinusoids}
\end{figure}

\vspace{-4pt}
\subsection{Difficulty of DNN-based phase reconstruction}
\label{sec:investigation}
\vspace{-2pt}

The first problem is the wrapping effect of phase.
Phase is given by the complex argument of a complex-valued spectrogram, and it is wrapped in $ [-\pi, \pi)$.
Hence, phase becomes discontinuous at $\pm \pi$ even when it rotates smoothly.
Training of a DNN for phase reconstruction should take this wrapping effect into account, but ordinary loss functions for regression problems, including MSE, do not consider such periodic nature of phase.

The second problem is the sensitivity to a shift of a waveform.
Considering the Fourier transform, its amplitude is shift-invariant, i.e., the amplitude does not change by shifting a waveform.
In contrast, its phase is sensitive to the shift, and thus an infinite number of Fourier phases can be considered for an amplitude spectrum.
Here, we experimentally show the same tendency of STFT amplitude and phase.
As an example, we consider an utterance from JSUT corpus \cite{jsut} and that shifted $0.5$ ms (8 samples).
Their STFT amplitude, phase, and time-directional phase derivative, i.e., instantaneous frequency, were shown in Fig.~\ref{fig: sinusoids}.
While their amplitudes look similar, phases were different as observed in the rightmost column of Fig.~\ref{fig: sinusoids}.
On the other hand, the instantaneous frequencies were almost the same as in the amplitude case.
According to this result, a DNN for direct phase estimation should learn a map reflecting small changes in amplitude spectrograms onto large differences in phases.
In general, training of a DNN as such an unstable map is not an easy task.
In contrast, it can be expected the estimation of phase derivatives is easier than that of phase for DNNs.

\vspace{-4pt}
\subsection{Related works}
\label{sec:takamichi}
\vspace{-2pt}

To address the first problem, \cite{takamichi} proposed a loss function based on the von Mises distribution.
The von Mises distribution is a distribution of a continuous periodic variable given by $p(\phi) = \mathrm{e}^{\kappa \cos(\phi - \mu)}/{2 \pi I_0(\kappa)}$,
where $\mu$ is a circular mean, $\kappa$ is a concentration, and $I_0(\kappa)$ is the modified Bessel function of the first kind of order $0$.
Its negative log-likelihood is given by
\begin{equation}
-\log p(\phi) = - \kappa \cos (\phi - \mu) + C(\kappa),
\label{eq: neglogvon}
\end{equation}
where $C(\kappa)$ is independent of $\mu$.
Derived from Eq.~\eqref{eq: neglogvon}, the loss function proposed in \cite{takamichi} is formulated by
\begin{equation}
\mathcal{L}(\theta) = - \sum_{\omega, \tau} \cos \bigl( \Phi_{\omega, \tau}^\star - \mathcal{F}_\text{Ph} (\boldsymbol{\psi}_\tau, \theta)_{\omega} \bigr),
\label{eq: vonloss}
\vspace{-4pt}
\end{equation}
where $\Phi_{\omega, \tau}^\star \in [-\pi, \pi )$ is the clean phase, and $\tau = 0, \ldots, T-1$ and $\omega = 0, \ldots, K-1$ are the time and frequency indices, respectively.
$\mathcal{F}_\text{Ph}$ is a DNN for direct phase estimation, $\theta$ is a set of parameters of the DNN, and $\boldsymbol{\psi}_\tau$ is the input feature calculated from an amplitude spectrogram for estimating phase at the $\tau$th time-frame.
This loss function is insensitive to the ambiguity of $2\pi$ of the estimated phase.

In addition, \cite{takamichi} also proposed the group delay loss, evaluating the group delay calculated from the estimated phase, because group delay has a strong relation to amplitude spectrogram.
Recently, some studies for DNN-based audio synthesis focused on phase derivatives \cite{arie,gansynth}.
To learn a generative model of complex-valued spectrograms, \cite{arie} uses both instantaneous frequency and group delay in its loss functions as an extension of the group delay loss.
On the other hand, \cite{gansynth} reconstructs phase by integrating the estimated instantaneous frequency along with the time-direction.

\vspace{-4pt}
\section{Proposed phase reconstruction}
\vspace{-4pt}

In the previous section, we describe the sensitivity of the phase to a waveform shift which makes the map for direct phase estimation unstable.
To overcome this problem, we propose a two-stage phase reconstruction method which uses the phase derivatives estimated by DNNs.
The overall procedure of the proposed phase reconstruction is illustrated in Fig.~\ref{fig: illust}.
In the proposed method, two DNNs estimate the instantaneous frequency and group delay to avoid the sensitivity problem.
Then, phase is recursively reconstructed by solving the least squares problem from its derivatives.

\vspace{-4pt}
\subsection{Estimation of phase derivatives}
\vspace{-2pt}
\label{sec: train}

In our proposed method, the instantaneous frequency and group delay at the $\tau$th time-frame are estimated by DNNs as follows:
\begin{align}
\mathbf{v}_{\tau} &= \mathcal{F}_\text{IF}(\boldsymbol{\psi}_{\tau}, \theta_\text{IF}), \label{eq:ifest} \\
\mathbf{u}_{\tau} &= \mathcal{F}_\text{GD}(\boldsymbol{\psi}_\tau, \theta_\text{GD}), \label{eq:gdest}
\vspace{-4pt}
\end{align}
where $\mathbf{v}_{\tau} = [v_{0,\tau}, \ldots, v_{K-1,\tau}]^\mathsf{T}$ is the estimated instantaneous frequency, and $\mathbf{u}_{\tau} = [u_{0,\tau}, \ldots, u_{K-2,\tau}]^\mathsf{T}$ is the estimated group delay.
DNNs are trained to minimize following loss functions:
\begin{align}
\mathcal{L}_\text{IF}(\theta_\text{IF}) &= -\sum_{\omega,\tau} \cos \bigl({V}^\star_{\omega,\tau} - \mathcal{F}_\text{IF}(\boldsymbol{\psi}_\tau, \theta_\text{IF})_{\omega}\bigr), \label{eq:ifloss}\\
\mathcal{L}_\text{GD}(\theta_\text{GD}) &= -\sum_{\omega,\tau} \cos \bigl(U^\star_{\omega,\tau} - \mathcal{F}_\text{GD}(\boldsymbol{\psi}_\tau, \theta_\text{GD})_{\omega} \bigr), \label{eq:gdloss}
\vspace{-6pt}
\end{align}
where ${V}^\star_{\omega,\tau} = \Phi_{\omega, \tau+1}^\star - \Phi_{\omega, \tau}^\star$ and ${U}^\star_{\omega,\tau} = - \Phi_{\omega+1, \tau}^\star + \Phi_{\omega, \tau}^\star$ are the instantaneous frequency and group delay%
\footnote{
We call the time- and negative frequency-directional phase derivatives as the instantaneous frequency and group delay for intuitiveness.
}%
defined by the numerical difference of the clean phase $\boldsymbol{\Phi}^\star$.
Phase derivatives are treated as periodic variables because they are approximated by the difference of wrapped phase.
Hence, estimated phase derivatives also have the ambiguity of $2\pi$, which is dealt in the next subsection.

\vspace{0pt}
\subsection{Recurrent phase unwrapping (RPU)}
\label{sec: rpu}
\vspace{-2pt}

\begin{figure}[t]
\centering
\includegraphics[width=0.99\columnwidth]{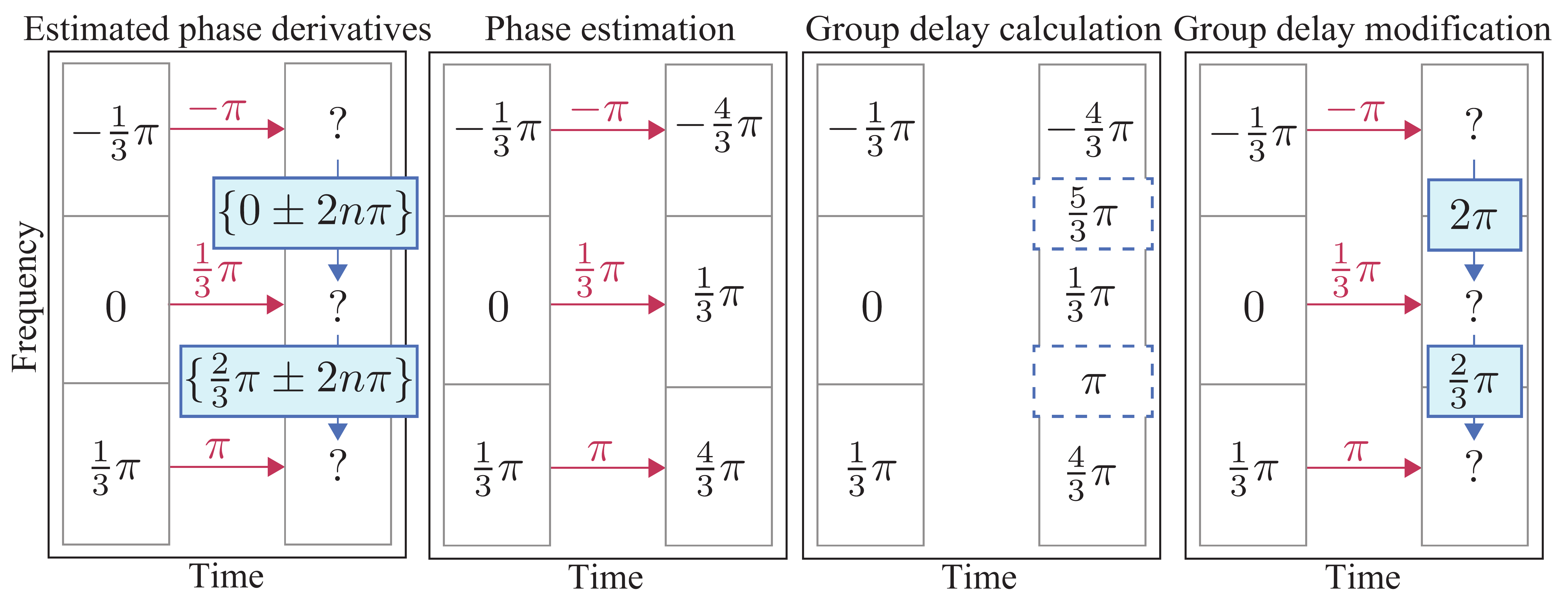}
\vspace{-4pt}
\caption{Illustration of ambiguity of estimated phase derivatives and its solution.
The red and blue arrows correspond to the instantaneous frequency and group delay, respectively.
In the second left figure, phase in the target time-frame is temporally estimated by Eq.~\eqref{eq:tempest}.
Then, the group delay is also calculated from the estimated phase as $\mathcal{D}_\omega(\hat{\phi}_\tau)$.
The rightmost figure solves the ambiguity of the group delay estimated by the DNN as in Eq.~\eqref{eq:gdmod}.}
\label{fig:ambiguity}
\vspace{-2pt}
\end{figure}

In this subsection, we explain the later part of the proposed phase reconstruction, i.e., recursive phase estimation from its derivatives, which is named as RPU.
We first explain a popular 2D phase unwrapping algorithm because RPU borrows an idea from it.
The estimation of unwrapped 2D phase from its derivatives is not obvious because the unwrapped phase should be consistent in both directions. 
Let $\mathbf{V}$ and $\mathbf{U}$ be given phase derivatives in each direction.
One of the famous methods for 2D phase unwrapping is formulated by the following least squares problem \cite{2dpu}:
\begin{equation}
\text{Find} \,\,\, {\boldsymbol{\Phi}} \in \text{arg}\min_{{\boldsymbol{\Phi}}} \| \mathcal{D}_\tau({\boldsymbol{\Phi}}) -  \mathbf{V} \|_F^2 + \| \mathcal{D}_\omega({\boldsymbol{\Phi}}) -  \mathbf{U} \|_F^2,
\label{eq: pu2d}
\end{equation}
where ${\boldsymbol{\Phi}}$ is the unwrapped phase, $\|\cdot\|_F$ is the Frobenius norm, $[\mathcal{D}_\tau({\boldsymbol{\Phi}})]_{\omega,\tau} = \Phi_{\omega,\tau+1} - \Phi_{\omega,\tau}$ is the time-directional difference operator, and $[\mathcal{D}_\omega({\boldsymbol{\Phi}})]_{\omega,\tau} = -\Phi_{\omega+1,\tau} + \Phi_{\omega,\tau}$ is the frequency-directional difference operators.
This method can be applied to phase reconstruction of a spectrogram by considering phase reconstruction as the estimation of unwrapped 2D phase.
However, this approach has the following two problems.
First, it requires huge computation for calculating the inverse of $KT \times KT$ matrix.
Second, the phase derivatives estimated the DNNs have the ambiguity of $2\pi$, which should be properly solved to maintain spatial consistency.

For avoiding the first problem, we propose RPU that recursively solves the following small least squares problem:
\begin{equation}
{\boldsymbol{\phi}}_\tau = \text{arg}\min_{{\boldsymbol{\phi}}} \| {\boldsymbol{\phi}} - {\boldsymbol{\phi}}_{\tau-1}^{\mathcal{W}} -  \mathbf{v}_{\tau-1} \|_2^2 + \| \mathcal{D}_\omega({\boldsymbol{\phi}}) -  {\mathbf{u}}_\tau \|_2^2, \label{eq:rpu}
\end{equation}
where ${\boldsymbol{\phi}}_{\tau-1}^{\mathcal{W}} = \mathcal{W}({\boldsymbol{\phi}}_{\tau-1})$ is the estimated wrapped phase in the $(\tau-1)$th time-frame, $\|\cdot\|_2$ is the Euclidean norm, and  $\mathcal{W}(\cdot) = [\cdot+\pi]_\mathrm{mod2\pi} -\pi$ is the wrapping operator%
\footnote{
We applied the wrapping operator to the solution of the least squares problem given in Eq.~\eqref{eq:rpu} for preventing large values which cause numerical instability.
Note that this wrapping operation does not affect the reconstructed waveform. 
}.
This least squares problem aims to estimate the phase which is consistent with the phase derivatives estimated by DNNs.
The least squares problem given in Eq.~\eqref{eq:rpu} is solved as
\begin{equation}
{\boldsymbol{\phi}}_\tau = (\mathbf{I} + \mathcal{D}_\omega^\mathsf{T} \mathcal{D}_\omega)^{-1} ( {\boldsymbol{\phi}}_{\tau-1}^{\mathcal{W}} + \mathbf{v}_{\tau-1} +\mathcal{D}_\omega^\mathsf{T} \mathbf{u}_\tau ),
\label{eq: solv1}
\end{equation}
where $\mathbf{I}$ is the identity matrix.
In Eq.~\eqref{eq: solv1}, the required computation is reduced to only $K\times K$ matrix inversion.
This is because RPU focuses on phase in the successive time frames and treats the phase in the previous time-frame as a fixed value while Eq.~\eqref{eq: pu2d} considers all time frames at the same time.

The second problem is the ambiguity of the phase derivatives estimated by DNNs.
The estimated instantaneous frequency and group delay have the ambiguity of $2\pi$ as discussed in the previous subsection, and we must solve this ambiguity appropriately.
For this goal, we fix the instantaneous frequency, and select the appropriate group delay, which is consistent with the instantaneous frequency, from the set $\{u_{\omega,\tau}\pm2n\pi \}$, where $n \in \mathbb{N}$.
The proposed group delay modification is conducted by
\begin{align}
\hat{\boldsymbol{\phi}}_\tau  &= \boldsymbol{\phi}_{\tau-1}^{\mathcal{W}} + \mathbf{v}_{\tau-1}, \label{eq:tempest}\\
\tilde{\mathbf{u}}_{\tau} &= \mathcal{D}_\omega(\hat{\boldsymbol{\phi}}_\tau) + \mathcal{W}(\mathbf{u}_{\tau} -  \mathcal{D}_\omega(\hat{\boldsymbol{\phi}}_{\tau})). \label{eq:gdmod}
\end{align}
This modification process is illustrated in Fig.~\ref{fig:ambiguity}.
In Eq.~\eqref{eq:tempest}, phase in the target time-frame is temporally estimated as shown in the second left figure.
Then, the appropriate group delay, which is the nearest to $\mathcal{D}_\omega (\hat{\boldsymbol{\phi}}_\tau)$, is selected from the set $\{ u_{\omega,\tau} \pm 2n\pi\}$ by Eq.~\eqref{eq:gdmod} as shown in the rightmost figure.
After solving the ambiguity, $\tilde{\mathbf{u}}_{\tau}$ is substituted to ${\mathbf{u}}_{\tau}$ in Eq.~\eqref{eq:rpu}.

\vspace{-4pt}
\subsection{Summary of the proposed phase reconstruction}
\vspace{-2pt}

In the training, pairs of clean amplitude spectrogram and phase are calculated from utterances by STFT, and the instantaneous frequency and group delay are calculated by the numerical difference of the phase.
Two DNNs for estimating the phase derivatives are trained by the loss functions given in Eqs.~\eqref{eq:ifloss} and \eqref{eq:gdloss}.
Then, the proposed phase reconstruction method is summarized as follows:
{
\setlength{\leftmargini}{12pt}         
\begin{itemize}
	\setlength{\itemsep}{4pt}      	
	\setlength{\parskip}{0pt}      	
	\setlength{\itemindent}{0pt}   	
	\setlength{\labelsep}{5pt}     	
\item[1.] Estimate the instantaneous frequency in the previous time-frame $\mathbf{v}_{\tau-1}$ and group delay in the target time-frame $\mathbf{u}_{\tau}$ as in Eqs.~\eqref{eq:ifest} and \eqref{eq:gdest}, respectively.
\item[2.] Reconstruct phase by RPU:
\begin{itemize}
\item[2.1] Modify the estimated group delay as in Eqs.~\eqref{eq:tempest} and \eqref{eq:gdmod}.
\item[2.2] Estimate phase in the next time-frame by Eq.~\eqref{eq: solv1}.
\item[2.3] Repeat 2.1 and 2.2 for all time-frames.
\end{itemize}
\end{itemize}
}
\noindent
After reconstructing phase in all time-frames, we can obtain the complex-valued spectrogram and convert it to the time-domain.

\vspace{-4pt}
\section{Experiments}
\vspace{-4pt}

To validate the effectiveness of the proposed method, the quality of reconstructed speeches from given amplitude spectrograms was evaluated.
It was compared with the direct phase estimation by the von Mises DNN \cite{takamichi}.
In addition, we also investigated the performance of the phase reconstruction by integrating the estimated instantaneous frequency as in \cite{gansynth}%
\footnote{
This method corresponds to reconstruct phase by Eq.~\eqref{eq:tempest}, which does not consider the group delay.
In contrast, the proposed method uses not only the instantaneous frequency but also group delay in RPU.
}%
.

\vspace{-4pt}
\subsection{Experimental condition}
\vspace{-2pt}

Evaluations were performed using the subsets of JSUT corpus \cite{jsut} (BASIC5000 and ONOMATOPE300) which is a Japanese speech corpus uttered by one female speaker.
Utterances from BASIC5000 were used as a training set where all utterances were resampled at $16$ kHz.
STFT was implemented with the Hann window, whose duration was $32$ ms with $8$ ms shift.
During the training, the utterances were divided into about 2-second-long segments.
In this experiment, the DNNs were trained by the Adam optimizer for $10,000$ epochs where the learning rate was decayed every $1,000$ epoch by multiplying $1/2$.
The initial learning rate was set to $0.01$.
In the testing, $300$ utterances from ONOMATOPE300 were used.

The input feature $\boldsymbol{\psi}_\tau$ was the vector which consists of the log-amplitude spectrogram at current and $\pm2$ frames with a normalization.
Fully connected DNNs with $4$ layers were used where each layer had $1024$ units, and gated $\tanh$ nonlinearity \cite{wavenet} was used for the activation except the last layer.

{\renewcommand\arraystretch{1.05}
\begin{table}[t!]
\centering
\footnotesize
\caption{Accuracy of the DNN estimations. Range of the accuracy is from $-1$ to $1$, and higher is better.}
\label{tab: acc}
\begin{tabular}{c|ccc}
\hline
& Phase & Instantaneous frequency & Group delay \\
\hline
\multirow{1}{*}{Training} & 0.238 & 0.675 & 0.702 \\
\hline
\multirow{1}{*}{Testing} & 0.003 & 0.644  & 0.646 \\
\hline
\end{tabular}
\vspace{-2pt}
\end{table}
}

\begin{figure}[t]
\centering
\includegraphics[width=0.99\columnwidth]{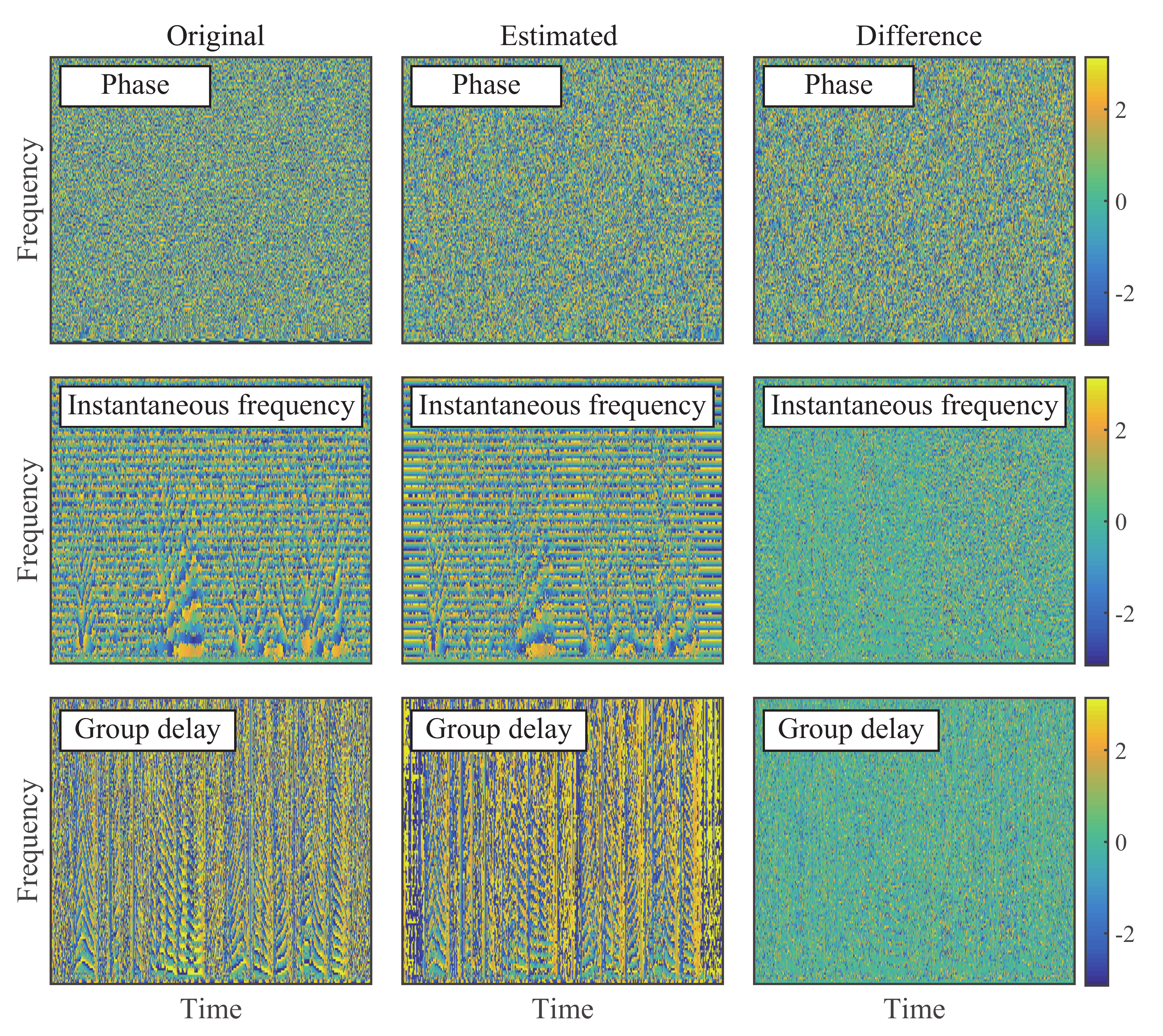}
\vspace{-2pt}
\caption{Examples of the estimated phase and its derivatives.
}
\label{fig: exam}
\vspace{-2pt}
\end{figure}

\begin{figure}[t]
\centering
\includegraphics[width=0.99\columnwidth]{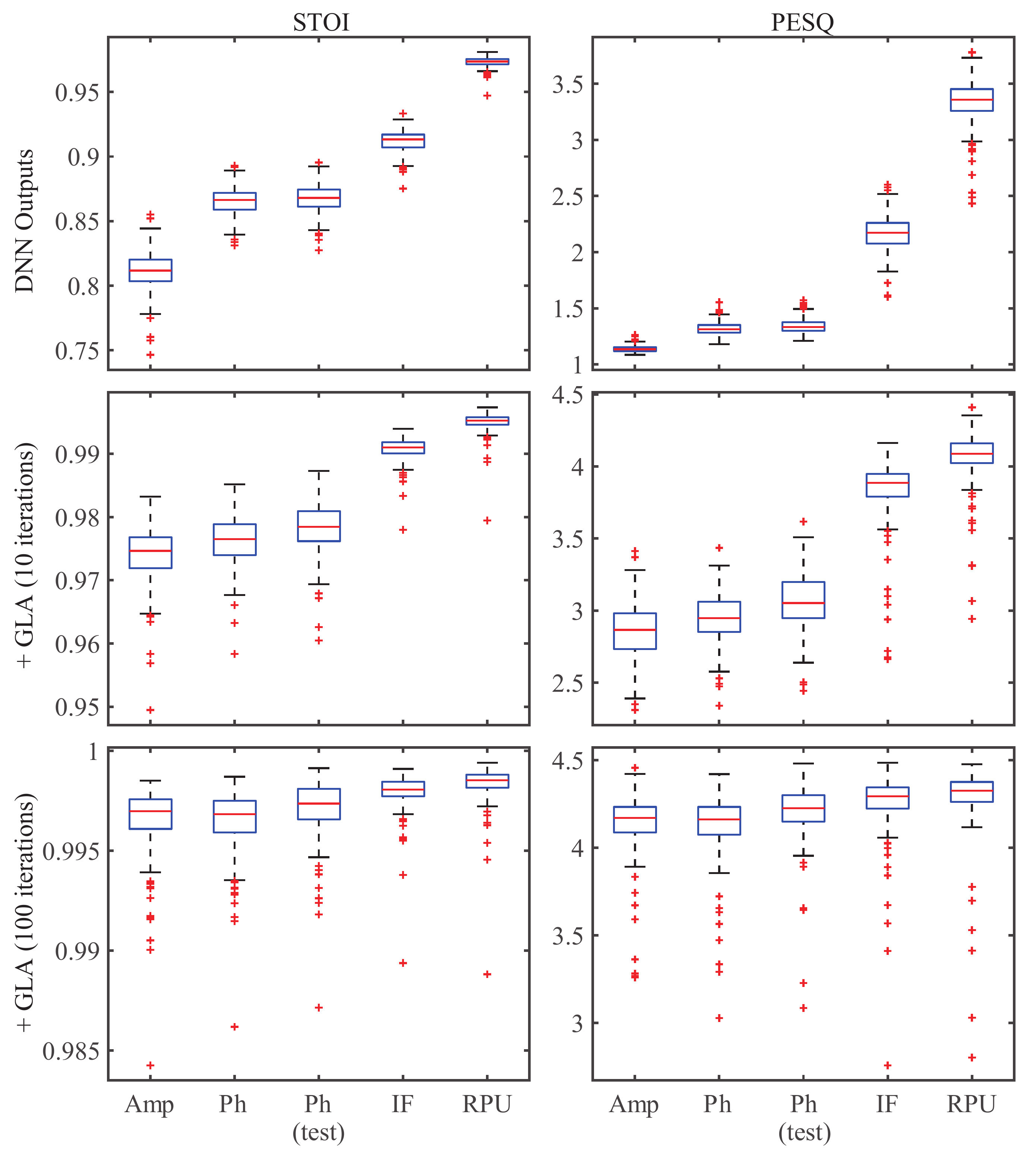}
\vspace{-6pt}
\caption{Boxplots of STOI and PESQ for $300$ reconstructed utterances from amplitude spectrograms. The boxes indicate the first and third quartiles. The top row shows the measures of the reconstructed speech by each method.
In the center and bottom rows, GLA was applied as post-processing $10$ and $100$ times, respectively.
}
\label{fig: evals}
\vspace{-4pt}
\end{figure}

\vspace{-4pt}
\subsection{Experimental results}
\vspace{-2pt}

We first evaluated the prediction accuracy of phase and its derivatives by DNNs.
Table~\ref{tab: acc} shows the accuracy of the estimation defined by $\sum_{\omega, \tau}\cos(\phi^\star_{\omega, \tau} - {\phi}_{\omega, \tau})/KT$ which takes a value in $[-1, 1]$.
The accuracy of the phase derivatives was higher than that of the phase itself in both training and testing.
This indicates that the estimation of phase derivatives are easier than that of phase itself as our expectation.
Considering phase derivatives, the difference between the accuracy in training and testing are relatively small, but the accuracy of phase estimation significantly decreased in testing.
That is, the DNN for direct phase estimation caused overfitting in our experimental condition.
We also trained the same DNN for direct phase estimation using the testing dataset (ONOMATOPE300).
Its accuracy was $0.557$ which was higher than the accuracy in the training, because the DNN overfitted to the small testing dataset. 
However, it was still lower than the accuracy of the phase derivatives in Table~\ref{tab: acc}.
Figure~\ref{fig: exam} shows examples of the phase and its derivatives estimated by DNNs.
We can observe the same structure in both clean and estimated phase derivatives.
On the other hand, there exists a large estimation error in phase, which is consistent with the quantitative evaluation in Table~\ref{tab: acc}.

The boxplots of  STOI \cite{stoi} and PESQ \cite{pesq} of the reconstructed signals are illustrated in Fig.~\ref{fig: evals}, where the direct phase estimation \cite{takamichi} and the numerical integration of the instantaneous frequency \cite{gansynth} are abbreviated as ``Ph'' and ``IF'', respectively.
Since the direct phase estimation caused overfitting as shown in Table~\ref{tab: acc}, we also evaluated the direct phase estimation trained by the testing dataset (ONOMATOPE300) as ``Ph (test)'' to approximately evaluate its best possible performance on the testing datasett.
We evaluated the zero-phase, i.e., without phase reconstruction, as ``Amp.''
In addition, we applied the Griffin--Lim algorithm (GLA) to each estimation \cite{gla}.
The proposed method outperformed the direct phase estimation by the von Mises DNNs and the numerical integration of the estimated instantaneous frequency with and without GLA.
Comparing to those conventional methods, the effectiveness of the proposed method was confirmed by one-side $t$-test ($p<0.01$) in both objective measures.
This result indicates the effectiveness of the phase reconstruction from both instantaneous frequency and group delay by RPU.
Our audio samples are available online%
\footnote{
\renewcommand{\UrlFont}{\ttfamily\scriptsize}
{\url{https://sites.google.com/view/yoshiki-masuyama/rpu}}
}%
.

\vspace{-4pt}
\section{Conclusion}
\vspace{-2pt}

In this paper, we proposed a DNN-based two-stage phase reconstruction method.
For reconstructing phase from instantaneous frequency and group delay estimated by DNNs, we proposed RPU which recursively reconstructs phase by solving the least squares problem.
The effectiveness of the proposed method was confirmed by comparing with the direct phase estimation by the von Mises DNN and phase reconstruction solely from the estimated instantaneous frequency.

\bibliographystyle{IEEEbib}

\begin{thebibliography}{10}

\bibitem{mowlaee}
P.~{Mowlaee} and J.~{Kulmer},
\newblock ``Phase estimation in single-channel speech enhancement:
  {L}imits-potential,''
\newblock {\em IEEE/ACM Trans. Audio Speech Lang. Process.}, vol. 23, no. 8,
  pp. 1283--1294, Aug. 2015.

\bibitem{audvis}
T.~Afouras, J.~Son Chung, and A.~Zisserman,
\newblock ``The conversation: {D}eep audio-visual speech enhancement,''
\newblock in {\em INTERSPEECH}, Sept. 2018, pp. 3244--3248.

\bibitem{taco}
Y.~Wang and et. al.,
\newblock ``Tacotron: Towards end-to-end speech synthesis,''
\newblock in {\em INTERSPEECH}, Aug. 2017, pp. 4006--4010.

\bibitem{takaki}
S.~Takaki, H.~Kameoka, and J.~Yamagishi,
\newblock ``Direct modeling of frequency spectra and waveform generation based
  on phase recovery for {D}{N}{N}-based speech synthesis,''
\newblock in {\em INTERSPEECH}, Aug. 2017, pp. 1128--1132.

\bibitem{saito}
Y.~Saito, S.~Takamichi, and H.~Saruwatari,
\newblock ``Text-to-speech synthesis using {S}{T}{F}{T} spectra based on
  low-/multi-resolution generative adversarial networks,''
\newblock in {\em IEEE Int. Conf. Acoust., Speech, Signal Process. (ICASSP)},
  Apr. 2018, pp. 5299--5303.

\bibitem{tfgan}
A.~Marafioti, N.~Holighaus, N.~Perraudin, and P.~Majdak,
\newblock ``Adversarial generation of time-frequency features with application
  in audio synthesis,''
\newblock {\em arXiv:1902.04072}, 2019.

\bibitem{melnet}
S.~Vasquez and M.~Lewis,
\newblock ``{M}el{N}et: {A} generative model for audio in the frequency
  domain,''
\newblock {\em ar{X}iv preprint ar{X}iv:1906.01083}, 2019.

\bibitem{phase1}
T.~Gerkmann, M.~Krawczyk-Becker, and J.~Le~Roux,
\newblock ``Phase processing for single-channel speech enhancement: {H}istory
  and recent advances,''
\newblock {\em IEEE Signal Process. Mag.}, vol. 32, no. 2, pp. 55--66, Mar.
  2015.

\bibitem{phase2}
P.~Mowlaee, R.~Saeidi, and Y.~Stylianou,
\newblock ``Advances in phase-aware signal processing in speech
  communication,''
\newblock {\em Speech Commun.}, vol. 81, pp. 1--29, July 2016.

\bibitem{complexmask}
D.~S. {Williamson} and D.~{Wang},
\newblock ``Time-frequency masking in the complex domain for speech
  dereverberation and denoising,''
\newblock {\em IEEE/ACM Trans. Audio, Speech, Lang. Process.}, vol. 25, no. 7,
  pp. 1, July 2017.

\bibitem{gla}
D.~Griffin and J.~Lim,
\newblock ``Signal estimation from modified short-time {F}ourier transform,''
\newblock {\em IEEE Trans. Acoust., Speech, Signal Process.}, vol. 32, no. 2,
  pp. 236--243, Apr. 1984.

\bibitem{fgla}
N.~Perraudin, P.~Balazs, and P.~L. S{\o}ndergaard,
\newblock ``A fast {G}riffin--{L}im algorithm,''
\newblock in {\em IEEE Workshop Appl. Signal Process. Audio Acoust.}, Oct.
  2013, pp. 1--4.

\bibitem{glaadmm}
Y.~Masuyama, K.~Yatabe, and Y.~Oikawa,
\newblock ``{G}riffin--{L}im like phase recovery via alternating direction
  method of multipliers,''
\newblock {\em IEEE Signal Process. Lett.}, vol. 26, no. 1, pp. 184--188, Jan.
  2019.

\bibitem{spsi}
G.~T. Beauregard, M.~Harish, and L.~Wyse,
\newblock ``Single pass spectrogram inversion,''
\newblock in {\em IEEE Int. Conf. Digit. Signal Process. (DSP)}, July 2015, pp.
  427--431.

\bibitem{consistency}
J.~{{Le} Roux}, N.~Ono, and S.~Sagayama,
\newblock ``Explicit consistency constraints for {S}{T}{F}{T} spectrograms and
  their application to phase reconstruction,''
\newblock in {\em ISCA Workshop Stat. Percept. Audit. (SAPA)}, Sept. 2008, pp.
  23--28.

\bibitem{stftpi}
M.~Krawczyk and T.~Gerkmann,
\newblock ``{S}{T}{F}{T} phase reconstruction in voiced speech for an improved
  single-channel speech enhancement,''
\newblock {\em IEEE/ACM Trans. Audio, Speech, Lang. Process.}, vol. 22, no. 12,
  pp. 1931--1940, Dec. 2014.

\bibitem{ipclr}
Y.~{Masuyama}, K.~{Yatabe}, and Y.~{Oikawa},
\newblock ``Low-rankness of complex-valued spectrogram and its application to
  phase-aware audio processing,''
\newblock in {\em IEEE Int. Conf. Acoust., Speech, Signal Process. (ICASSP)},
  May 2019, pp. 855--859.

\bibitem{unfold}
Z.-Q. Wang, J.~{Le Roux}, D.~Wang, and J.~Hershey,
\newblock ``End-to-end speech separation with unfolded iterative phase
  reconstruction,''
\newblock in {\em INTERSPEECH}, Sept. 2019, pp. 2708--2712.

\bibitem{oyamada}
K.~Oyamada, H.~Kameoka, T.~Kaneko, K.~Tanaka, N.~Hojo, and H.~Ando,
\newblock ``Generative adversarial network-based approach to signal
  reconstruction from magnitude spectrograms,''
\newblock in {\em Eur. Signal Process. Conf. (EUSIPCO)}, Sept. 2018.

\bibitem{degli}
Y.~Masuyama, K.~Yatabe, Y.~Koizumi, Y.~Oikawa, and N.~Harada,
\newblock ``Deep {G}riffin--{L}im iteration,''
\newblock in {\em IEEE Int. Conf. Acoust., Speech, Signal Process. (ICASSP)},
  May 2019, pp. 61--65.

\bibitem{pnet}
N.~Takahashi, P.~Agrawal, N.~Goswami, and Y.~Mitsufuji,
\newblock ``{P}hase{N}et: {D}iscretized phase modeling with deep neural
  networks for audio source separation,''
\newblock in {\em INTERSPEECH}, 2018.

\bibitem{pbook}
J.~{Le}. Roux, G.~Wichern, S.~Watanabe, A.~Sarroff, and J.~R. Hershey,
\newblock ``Phasebook and friends: Leveraging discrete representations for
  source separation,''
\newblock {\em IEEE J. Sel. Top. Signal Process.}, 2019.

\bibitem{takamichi}
S.~Takamichi, Y.~Saito, N.~Takamune, D.~Kitamura, and H.Saruwatari,
\newblock ``Phase reconstruction from amplitude spectrograms based on
  von--{M}ises-distribution deep neural network,''
\newblock in {\em Int. Workshop Acoust. Signal Enhance. (IWAENC)}, Sept. 2018,
  pp. 286--290.

\bibitem{flandrin}
F.~Auger, {\'E}.~Chassande-Mottin, and P.~Flandrin,
\newblock ``On phase-magnitude relationships in the short-time {F}ourier
  transform,''
\newblock {\em IEEE Signal Process. Lett.}, vol. 19, no. 5, pp. 267--270, May
  2012.

\bibitem{prusa}
Z.~Pr{\r u}\v{s}a, P.~Balazs, and P.~L. S{\o}ndergaard,
\newblock ``A noniterative method for reconstruction of phase from {S}{T}{F}{T}
  magnitude,''
\newblock {\em IEEE/ACM Trans. Audio, Speech, Lang. Process.}, vol. 25, no. 5,
  pp. 1154--1164, May 2017.

\bibitem{ast}
K.~Yatabe, Y.~Masuyama, T.~Kusano, and Y.~Oikawa,
\newblock ``Representation of complex spectrogram via phase conversion,''
\newblock {\em Acoust. Sci. \& Tech.}, vol. 40, no. 3, pp. 170--177, May 2019.

\bibitem{2dpu}
D.~C. Ghiglia and M.~D. Pritt,
\newblock {\em Two-dimensional phase unwrapping: theory, algorithms, and
  software},
\newblock Wiley, 1998.

\bibitem{jsut}
R.~Sonobe and S.~Takamichi,
\newblock ``{J}{S}{U}{T} corpus: free large-scale japanese speech corpus for
  end-to-end speech synthesis,''
\newblock {\em arXiv:1711.00354}, 2017.

\bibitem{arie}
A.~A. {Nugraha}, K.~{Sekiguchi}, and K.~{Yoshii},
\newblock ``A deep generative model of speech complex spectrograms,''
\newblock in {\em IEEE Int. Conf. Acoust., Speech, Signal Process. (ICASSP)},
  May 2019, pp. 905--909.

\bibitem{gansynth}
J.~Engel, K.~K. Agrawal, S.~Chen, I.~Gulrajani, C.~Donahue, and A.~Roberts,
\newblock ``{G}{A}{N}{S}ynth: {A}dversarial neural audio synthesis,''
\newblock in {\em Int. Conf. Learn. Represent. (ICLR)}, 2019.

\bibitem{wavenet}
A.~Oord and et. al.,
\newblock ``Wavenet: {A} generative model for raw audio,''
\newblock {\em ar{X}iv preprint ar{X}iv:1609.03499}, 2016.

\bibitem{stoi}
C.~H. Taal, R.~C. Hendriks, R.~Heusdens, and J.~Jensen,
\newblock ``An algorithm for intelligibility prediction of time-frequency
  weighted noisy speech,''
\newblock {\em IEEE Trans. Audio, Speech, Lang. Process,}, vol. 19, no. 7, pp.
  2155--2136, 2011.

\bibitem{pesq}
{\em P.862.2: Wideband extension to Recommendation P.862 for the assessment of
  wideband telephone networks and speech codecs}, ITU-T Std. P.862.2, 2007.

\end{thebibliography}

\end{document}